\documentclass{article}

\PassOptionsToPackage{numbers, compress}{natbib}

\usepackage[final]{neurips_2021}

\usepackage[utf8]{inputenc} 
\usepackage[T1]{fontenc}    
\usepackage{hyperref}       
\usepackage{url}            
\usepackage{booktabs}       
\usepackage{graphicx}
\usepackage{amsfonts}       
\usepackage{nicefrac}       
\usepackage{microtype}      
\usepackage{xcolor}         

\title{Classification of histopathology images using ConvNets to detect Lupus Nephritis}

\author{%
  Akash Gupta\thanks{Work done partially during an internship at IIIT Hyderabad}\\
  New York University\\
  \texttt{aksg@nyu.edu} \\
  \And
  Anirudh Reddy \\
  IIIT Hyderabad  \\
  \texttt{anirudhreddy.r@research.iiit.ac.in} \\
  \AND
  CV Jawahar \\
  IIIT Hyderabad  \\
  \texttt{jawahar@iiit.ac.in}
  \And
  PK Vinod \\
  IIIT Hyderabad  \\
  \texttt{vinod.pk@iiit.ac.in} \\
}

\begin{document}

\maketitle

\begin{abstract}
  Systemic lupus erythematosus (SLE) is an autoimmune disease in which the immune system of the patient starts attacking healthy tissues of the body. Lupus Nephritis (LN) refers to the inflammation of kidney tissues resulting in renal failure due to these attacks. The International Society of Nephrology/Renal Pathology Society (ISN/RPS) has released a classification system based on various patterns observed during renal injury in SLE \cite{Weening241}. Traditional methods require meticulous pathological assessment of the renal biopsy and are time-consuming. Recently, computational techniques have helped to alleviate this issue by using virtual microscopy or Whole Slide Imaging (WSI). With the use of deep learning and modern computer vision techniques, we propose a pipeline that is able to automate the process of 1) detection of various glomeruli patterns present in these whole slide images and 2) classification of each image using the extracted glomeruli features.
\end{abstract}

\section{Introduction}
Systemic Lupus erythematosus (SLE) is a multisystem autoimmune disease, in which the person's own immune system attacks and damages the healthy tissues and organs. SLE can attack a wide range of places in the body. Lupus Nephritis (LN) refers to the case of SLE affecting the kidney and if left unchecked, LN can progress to renal failure. The classification of lupus nephritis is critical to the issue of patient care and for the comparing results of therapeutic trials between different clinics. The International society of Nephrology/Renal Pathology Society(ISN/RPS) has released the classification of LN into 6 classes (Class I-VI) based on the presence of various glomeruli features seen in patients affected by LN \cite{Weening241}. The major diagnosis of LN is done by a renal biopsy; the observed histological changes within glomeruli guide treatment of patients and are correlated with patient outcome. Several histological stains such as PAS and H\&E reveal different information regarding the disease features. Digitising such stained slides into images helps in developing computational algorithms which aid to the pathologists during diagnosis.

\begin{figure*}[t]
\begin{center}
\centering
    \includegraphics[width=0.85\linewidth]{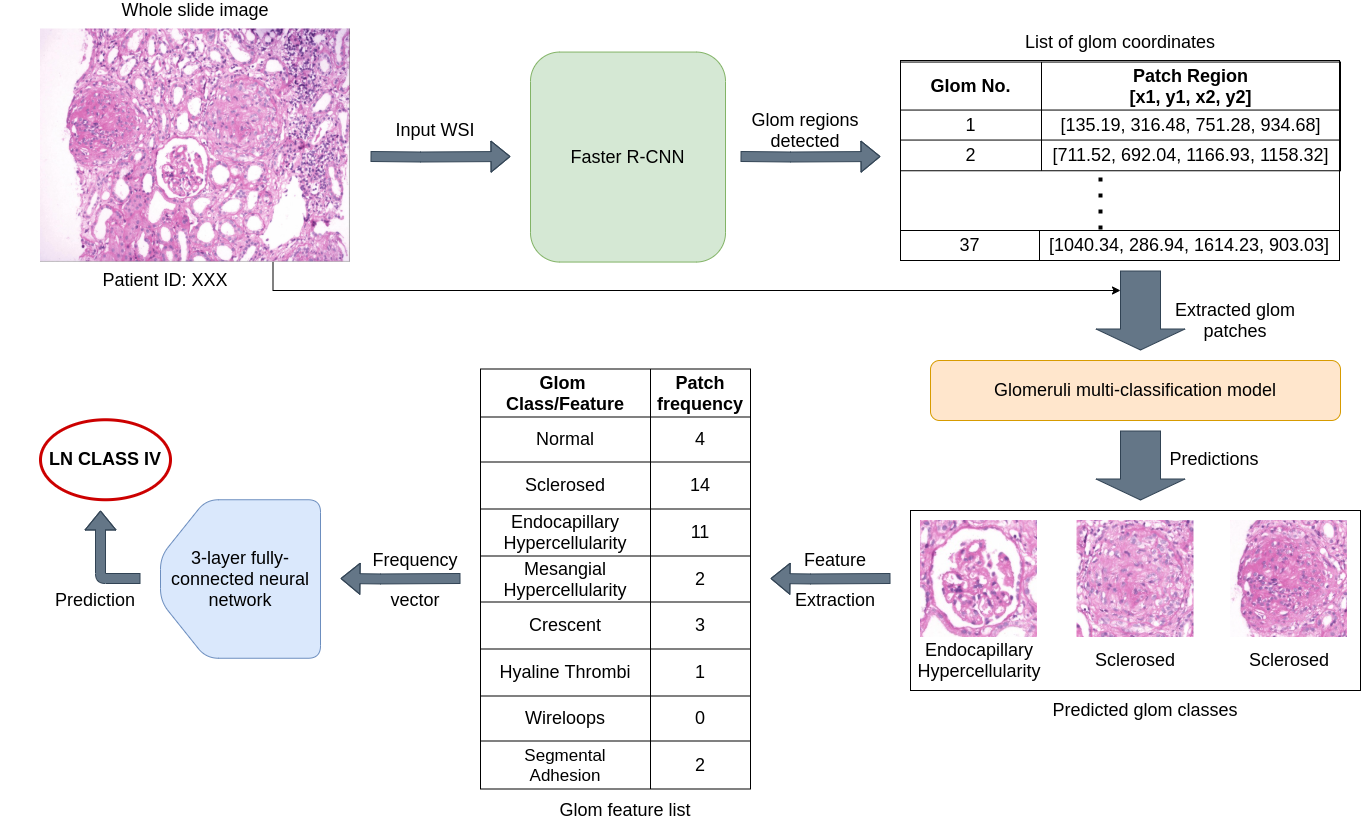}
\end{center}
   \caption{Glomeruli detection and classification pipeline}
\label{fig:short}
\end{figure*}

Typically the biopsy images are stored as Whole Slide Images(WSIs). A WSI is a digitized image of an entire microscopy slide and have huge resolution ($approx.\:80,000\times20,000$). Contrary to the numerous works published for glomeruli detection and classification \cite{electronics9030503, 10.1117/12.2295446, KANNAN2019955, Ginley1953}, only a few works exist that propose a complete pipeline for classification of WSIs into LN classes using glomeruli features. Additionally, they only consider limited number of glomeruli classes: sclerosed v/s normal \cite{BUENO2020105273} or glomerular v/s non-glomerular \cite{jimaging4010020} , which is insufficient for LN classification. \cite{9263102} focused on the glomeruli classification and kidney-level classification. They extracted the glomeruli manually and used a bulk labeling scheme in order to avoid manual annotation by pathologists. This technique adds label noise in the dataset as glomeruli images were given the same label as LN class regardless of their phenotype. They further used an approximate Bayesian DenseNet to resolve this issue claiming that DNNs are resistant to label noise. Although, they show significantly good results using this method, the scalability and genuinity from a pathologist's view seems unclear. As reported in \cite{Weening241}, it is important to consider the phenotypical features of glomeruli present in the kidney while in the classification of LN. In our method (shown in Figure \ref{fig:short}), we propose an end-to-end automated method from extracting glomeruli from the images to classifying them into various stages of Lupus Nephritis.

\section{Data collection}

 We collected biopsy samples from 60 patients diagnosed with various stages of LN from the Nizam Insitute of Medical Sciences (NIMS), Hyderabad, India. These are 108 WSIs recorded at different levels of magnification (10x, 20x and 40x). Due to unavailability of the image scanner, these images were taken by attaching a camera to the microscope and taking snapshots of the interesting regions of the biopsy tissue sample. The extracted regions were downsampled and saved as an image file of lower resolution ($approx\:2048\times1024$). The samples were stained by PAS (Periodic Acid Schiff) and color normalisation described by \cite{7460968} was performed. We used Aperio ImageScope\footnote{\url{https://www.leicabiosystems.com/digital-pathology/manage/aperio-imagescope}} software tool to extract and annotate the glomeruli from the provided images.  The labels of these glomeruli were validated by 2 pathologists from NIMS. Following the classification of LN proposed by the ISN/RPS, we divided the extracted glomeruli into 9 classes based on the glomerular features. In order to increase the dataset size, we used CircleMix augmentation \cite{lu2021improve}. Finally, we have 2280 glomeruli images with 9 classes as shown in Table \ref{tab:glomclass1}
 
\begin{table}[ht]
\begin{center}
\begin{tabular}{|l|c|}
\hline
Class & count \\
\hline\hline
Normal & 18 \\
Sclerosed & 25 \\
Endocapillary Hypercellularity & 19 \\
Messangial Hypercellularity & 10\\
ThickGBM & 41 \\
Wireloops & 10\\
Hyaline Trombi & 5 \\
Cresent & 14\\
Segmental Adhesion & 2\\
\hline
\end{tabular}
\end{center}
\caption{Glomeruli classes present in the WSI for a patient with LN stage IV in the NIMS dataset.}
\label{tab:glomclass1}
\end{table}

\section{Methods}
\subsection{Glomeruli detection}

Although there are many approaches towards segmenting glomeruli \cite{BUENO2020105273, jimaging4010020, electronics9030503, 10.1117/12.2295446, KANNAN2019955, Ginley1953}, they use very deep architectures for this task. We formulated the problem of extracting glomeruli as an object-detection task. We employed a region proposal based technique and used a Faster R-CNN model for locating bounding boxes for the glomeruli present in the images provided by NIMS. We used a VGGNet pretrained on ImageNet as the base network for generating the convolutional feature map. We fine-tuned the Faster R-CNN model on the slide images provided by NIMS for 50 epochs with a batch size of 5.

\subsection{Glomeruli Multi-Classification}

Based on the predicted coordinate values, we crop the glomeruli from the slide images and only take the true positives and further classified them into the 9 classes. Since a glomeruli image can have multiple labels, we used a multi-label multi-class classification approach. The model accepts the glomeruli image and passes it to 2 convolutional blocks. The resulting feature map is flattened to a feature vector and passed to a classifier with a sigmoid activation to output a vector of size $9\times1$ containing probabilities for each class. We found the best threshold value to be 0.45. If the probability of a class is greater than the threshold, we assign the class to the image.

\subsection{LN classification}

Further we aggregate all the glomeruli images based on their patient ID and generate a list of features. Once the glomeruli features list is obtained for each patient that depicts the types of glomeruli features present along with their counts, the final prediction of LN class of the patient is performed using a slide level approach. Here a frequency vector is generated for each patient based on the obtained glomeruli features list which is used as input to a 3-layer fully connected neural network to get the patient level LN classification.

\section{Results}

For glomeruli detection, out of the 108 images from NIMS, 100 were kept for training and 8 for test. The dataset was divided using patient ID such that slide images from the same patient are not found in both the training and validation sets. The results on the test set are shown in Table \ref{tab:detection}. For glomeruli multi-label classification, the model was pre-trained on \cite{BUENO2020105314} dataset for 30 epochs. This dataset contains 2340 glomeruli images extracted from 31 WSIs of PAS-stained kidney biopsies generated in European project AIDPATH. The dataset contains only 2 classes of glomeruli: normal (1170 images) and sclerosed (1170 images). We fine-tuned the model for another 20 epochs on our extracted set of 2280 glomeruli images from NIMS data where we divide the data into a 60-40\% split. To deal with label imbalance, weighted oversampling was used. We achieved an accuracy of 82.31\% with a precision of 0.782 and a recall of 0.87 on the test set. For LN classification, we achieved an accuracy of 83.2\% with a precision of 0.87 and a recall of 0.91. 
\begin{table}[ht]
\begin{center}
\begin{tabular}{|c|c|c|c|}
\hline
 & IoU$_{0.5:0.95}$ & IoU$_{0.5}$ & IoU$_{0.75}$ \\
\hline\hline
AP & 0.73 & 0.976 & 0.843 \\
AR & 0.742 & 0.792 & 0.792 \\
\hline
\end{tabular}
\end{center}
\caption{Object detection scores on NIMS validation set.}
\label{tab:detection}
\end{table}
\section{Conclusion}
We propose a complete pipeline for the classification of WSIs into various LN stages based on glomeruli features. The steps include 1) detection of glomeruli from PAS stained images, 2) multi-classification of these glomeruli into the 9 classes selected based on the glomeruli features seen in LN affected kidney biopsies and 3) predict the patient's stage of Lupus Nephritis (CLASS I to Class VI).  

\section{Broader Impact}

Our work has several useful implications. To our knowledge this forms the first work to use the ability of convolution neural networks for detecting Lupus Nephritis using visual glomerular features. Ginley et al. \cite{10.1117/12.2548528} shows the classification of 6 stages of LN by using hand-crafted glomerular features, but we believe that this limits the available feature set. Contrary to this, deep learning based techniques, allow the model to implicitly learn all the relevant features.  We would like to emphasize that our work does not intend to replace the decision-making of nephropathologists. Nevertheless, this work may have negative implications. It is worth noting that due to the limited availability of LN data we have only used a set of 9 commonly seen glomerular classes but in reality there are more than 25 glomerular classes that can be seen in some patients. Using images with such glomerular classes can alter the predictions and can lead to an increase in false positives. 

\section{Acknowledgement}

We thank Nizam Institute of Medical Sciences (NIMS), Hyderabad for the dataset and clinical guidance and IHub-Data, IIIT Hyderabad for financial support. 

{\small
\bibliographystyle{ieee}
\bibliography{neurips_2021}
}

\end{document}